%% file: main.tex
\documentclass[
	a4paper, 
	10pt, 
	unnumberedsections, 
	twoside, 
]{LTJournalArticle}

\runninghead{} 

\footertext{\textit{RISC-V Summit Europe, Barcelona, 5-9th June 2023}} 

\title{SafeTI Traffic Injector Enhancement for Effective Interference Testing in Critical Real-Time Systems}

\author{ 
  Francisco Fuentes\textsuperscript{1,2}, Raimon Casanova\textsuperscript{2}, Sergi Alcaide\textsuperscript{1} and Jaume Abella\textsuperscript{1}
}

\date{ 
  \footnotesize\textsuperscript{\textbf{1}}Barcelona Supercomputing Center (BSC), Barcelona, Spain\\
  \textsuperscript{\textbf{2}}Microelectronic and Electronic Systems Department, Universitat Aut\`{o}noma de Barcelona (UAB), Bellaterra, Spain
}

\renewcommand{\maketitlehookd}{%
	\begin{abstract}
    \noindent Safety-critical domains, such as automotive, space, and robotics, are adopting increasingly powerful multicores with abundant hardware shared resources for higher performance and efficiency. However, mutual interference due to parallel operation within the SoC must be properly validated. Recently, the SafeTI traffic injector has been released and integrated in a homogeneous RISC-V multicore for testing, otherwise untestable casuistic for software-only solutions. This paper introduces some enhancements performed on the SafeTI, which include internal pipelining for higher-rate traffic injection, and its tailoring to multiple interfaces, as well as its integration in a more powerful heterogeneous RISC-V multicore based on Gaisler’s technology for the space domain.
	\end{abstract}
}

\begin{document}

\maketitle

\input{1.0.introduction.tex}
\input{2.0.SafeTI.tex}
\input{3.0.Future.tex}

\input{4.0.Conclusions.tex}


\fontsize{8}{10}\printbibliography 

\end{document}

%% file: 1.0.introduction.tex
\section{Introduction}



Increasing performance demands in safety-critical real-time systems impose the adoption of Multi-Processor Systems-on-Chip (MPSoCs).
MPSoCs include multiple cores and accelerators that run several tasks in parallel, sharing hardware resources for efficiency reasons (e.g., on-chip caches, memory controllers, I/O interfaces).
Such sharing brings timing interference across cores, accelerators, and I/O interfaces, since several devices may contend for access to a specific device unable to serve requests from different sources simultaneously. Hence, requests are serialized, causing execution time delays on the affected tasks.\footnotetext{This work has been partially supported by the Spanish Ministry of Science and Innovation under grant PID2019-107255GB-C21 funded by MCIN/AEI/10.13039/501100011033.}

The development process of safety-critical systems is guided by domain-specific safety standards and guidelines (e.g., ISO26262 for automotive) that impose safety requirements to the system, which often include real-time requirements (e.g., braking the car within a specific timeframe since the braking pedal is activated). The architectural design of the system is devised to meet those safety requirements by construction, but the safety development process also imposes testing on the system to validate that safety requirements are effectively preserved.

In the case of timing interference, it is particularly challenging to test key performance corners, such as interference caused by asynchronous activity (e.g., due to traffic arriving through an Ethernet port),
especially if the type of transactions that should be tested cannot be induced synchronously by the cores, hence precluding the use of software tests to validate those cases.

Recently, the SafeTI traffic injector~\cite{SafeTI,SafeTIgit} has been proposed to tackle this gap. The SafeTI is capable of producing programmable traffic patterns, with high flexibility and controllability, hence allowing to test scenarios synchronously despite the fact they would only occur asynchronously in practice. Unfortunately, the current implementation of the SafeTI only works with AMBA Advanced High-performance Bus (AHB) interfaces, has been proven only in a 4-core homogeneous \mbox{RISC-V} multicore~\cite{SCC}, and has some limitations to inject traffic at a sufficiently high frequency.

This paper introduces the enhancements performed in the SafeTI to remove most of its constraints, such as pipelining its architecture for higher injection frequency, porting it to the popular AMBA Advanced eXtensible Interface (AXI), and integrating it in multiple interfaces of a more powerful heterogeneous MPSoC based on Frontgrade Gaisler technology for the space domain~\cite{SELENEgit}. We also provide some future prospects for this component.

%% file: 2.0.SafeTI.tex
\section{SafeTI Enhancements}
\label{sec:safeti}

\begin{figure}[t!]
\centering
  \includegraphics[width=0.9\columnwidth]{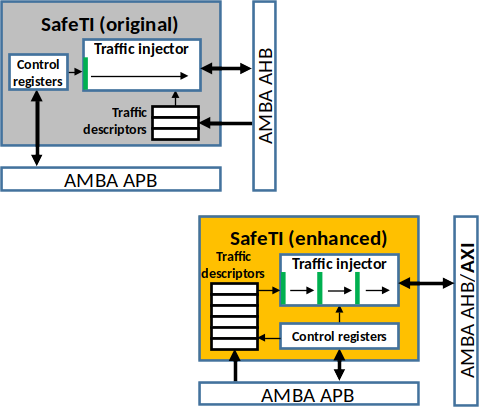}
  \caption{Original SafeTI architecture (top), and enhanced one (bottom).}
  \label{fig:safeti}
  \vspace{-0.4cm}
\end{figure}

\textbf{Overview}. The SafeTI is a programmable traffic injector consisting of a traffic descriptor buffer (see Figure~\ref{fig:safeti}, top), a set of control registers, and the traffic injector itself, which consumes the descriptors from the buffer as dictated by the control registers that, for instance, indicate when the traffic injection should start. 
Traffic descriptors encode the traffic pattern to be generated, including a target address, whether the access is read or write, the amount of data to transfer, and repetitions.
The original realization of the SafeTI has some limitations that we have removed. The remaining of this section describes the main ones.

\textbf{Initialization of the descriptors}. Originally, the SafeTI was connected to the AMBA AHB for both traffic injection and configuration (see Figure~\ref{fig:safeti}, top). However, programming descriptors in the buffer through the AMBA AHB data interface brought some concerns: (i) buffer updates could experience and cause timing interference, (ii) buffer updates could pollute cache memories, and (iii) buffer updates could become simply wrong.

Regarding timing interference, fetching descriptors through the same interface used for data transfers in the cores could create unforeseen interference in the cores.
Regarding cache pollution, since descriptors were fetched through the regular data path, they could pollute cache memories evicting useful contents.
Finally, if the SafeTI did not have access to coherent data from the cores, the descriptors
would not be available for the traffic injector to load, retrieving obsolete contents from memory instead of the correct traffic pattern.
To solve these limitations, we have enabled the use of the AMBA Advanced Peripheral Bus (APB) for configuration purposes. 
By being a separate interface, it neither interferes with data transfers, nor pollutes caches, nor experiences coherence problems since the APB interface writes coherent data to any SafeTI module. 
This is illustrated in Figure~\ref{fig:safeti}, bottom.

\textbf{Injection rate}. The original design of the SafeTI was not pipelined, which implied that some cycles were needed from the time a descriptor was fetched until the corresponding traffic was injected. This characteristic limited the injection rate
since back-to-back descriptors needed to be fetched and decoded prior to execution.
To solve this limitation, we have pipelined the design of the SafeTI, hence parallelizing descriptor fetch, decode, and execution and enabling sustained traffic limited by the interface only.

\begin{figure}[t!]
\vspace{-0.5cm}
\centering
  \includegraphics[width=0.95\columnwidth]{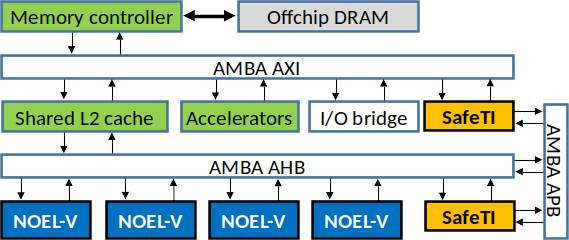}
  \caption{SELENE SoC with 2 SafeTI modules.}
  \label{fig:selenesoc}
  \vspace{-0.3cm}
\end{figure}

\textbf{Target interfaces}. Originally, the SafeTI was developed to support AMBA AHB only, since that was the protocol of the interface where it was first integrated~\cite{SafeTI,SCC}. However, the architecture of the SafeTI is not restricted to any particular interface and, in principle, could be tailored to operate with virtually any interface.
In our case, we have already extended it to work with AMBA AXI interfaces, such as those of the SELENE SoC~\cite{SELENEgit}, which is a RISC-V based MPSoC based on Gaisler's technology and intended
for domains such as space, railway, and automotive. This is illustrated in Figure~\ref{fig:selenesoc}, where we can see a schematic of the SELENE SoC with a SafeTI instance attached to the AHB interface and another to the AXI interface.

%% file: 3.0.Future.tex
\section{Future Plans}
\label{sec:future}

Our future prospects for the SafeTI include performing a more exhaustive verification and validation with the aim of easing its adoption,
and devising appropriate descriptors for an exhaustive test campaign of the SELENE SoC as an illustrative example of the use of the SafeTI. In the mid-term, we also aim at tailoring the descriptors and, potentially, the SafeTI design itself to enable new applications in the area of security (e.g., to counteract side-channel attacks) and testing (e.g., test cache coherence protocols).

%% file: 4.0.Conclusions.tex
\section{Summary}
\label{sec:concl}

The SafeTI is a powerful tool to test performance corners with limited effort, but its original design had a number of limitations. This work shows that those limitations have been removed, and the current design is more efficient and portable.